# On Real-Time Communication Systems with Noisy Feedback

Aditya Mahajan and Demosthenis Teneketzis

Department of EECS, Univ. of Michigan, Ann Arbor, MI – 48109. {adityam,teneket}@eecs.umich.edu

*Abstract*— We consider a real–time communication system with noisy feedback consisting of a Markov source, a forward and a backward discrete memoryless channels, and a receiver with finite memory. The objective is to design an optimal communication strategy (that is, encoding, decoding, and memory update strategies) to minimize the total expected distortion over a finite horizon. We present a sequential decomposition for the problem, which results in a set of nested optimality equations to determine optimal communication strategies. This provides a systematic methodology to determine globally optimal joint source–channel encoding and decoding strategies for real–time communication systems with noisy feedback.

## I. Problem Formulation

Consider a real–time communication system with noisy feedback as shown in Figure 1. This system consists of a source, a real–time encoder, a noisy forward channel, a noisy backward channel, and a real–time decoder with finite memory. The communication system operates in discrete time for a time horizon $T$.

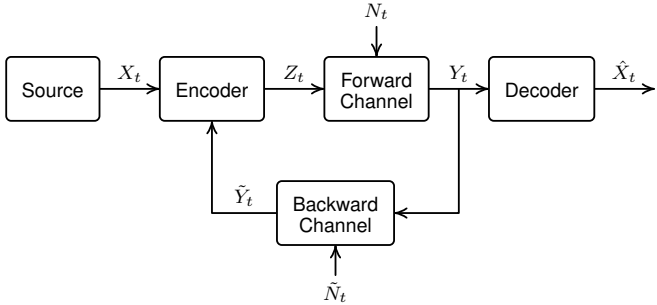

Fig 1. A real–time communication system with noisy feedback

At each stage $t$, the source produces an output $X_t$ taking values in a finite alphabet $\mathcal{X}$. We assume that the output sequence $\{X_t,\ t=1,\ldots,T\}$ forms a first–order Markov chain with initial distribution $P_{X_1}$ and matrix of transition probabilities $P_{X_{t+1}|X_t}$.

The communication system consists of two channels: a forward channel and a backward channel. We assume that both channels are independent DMC (discrete memoryless channels). The forward channel is a $|\mathcal{Z}|$–input $|\mathcal{Y}|$–output DMC, while the backward channel is a $|\mathcal{Y}|$–input $|\tilde{\mathcal{Y}}|$–output DMC. These channels can be described as

$$Y_t = h(Z_t, N_t), \qquad t = 1,\ldots,T, \tag{1a}$$
$$\tilde{Y}_{t-1} = \tilde{h}(Y_{t-1}, \tilde{N}_{t-1}), \quad t = 2,\ldots,T, \tag{1b}$$

where $h(\cdot)$ and $\tilde{h}(\cdot)$ denote the forward and backward channels at time $t$, respectively; $Z_t$ and $Y_{t-1}$ are the inputs to the forward and the backward channels at time $t$, respectively; $Y_t$ and $\tilde{Y}_{t-1}$ are the outputs of the forward and the backward channels at time $t$, respectively; and $N_t$ and $\tilde{N}_{t-1}$ are the channel noise in the forward and the backward channels at time $t$, respectively. The sequential order in which these system variables are generated is shown in Figure 2. The variables $Z_t$, $Y_t$, $\tilde{Y}_t$, $N_t$, and $\tilde{N}_t$ take values in finite alphabets $\mathcal{Z}$, $\mathcal{Y}$, $\tilde{\mathcal{Y}}$, $\mathcal{N}$, and $\tilde{\mathcal{N}}$, respectively. We assume that $\{N_t,\ t=1,\ldots,T\}$ and $\{\tilde{N}_t,\ t=1,\ldots,T\}$ are sequences of i.i.d. random variables with PMF (probability mass function) $P_N$ and $P_{\tilde{N}}$, respectively. These sequences are independent of each other and are also independent of the source output $\{X_t,\ t=1,\ldots,T\}$.

At each stage $t$, the encoder observes the output $X_t$ of the source and the output $\tilde{Y}_{t-1}$ of the backward channel. It generates an encoded symbol $Z_t$ using all its past observations using an encoding rule $c_t$, i.e.,

$$Z_1 = c_1(X_1), \tag{2a}$$
$$Z_t = c_t(X^t, Z^{t-1}, \tilde{Y}^{t-1}), \qquad t = 2,\ldots,T, \tag{2b}$$

where $X^t$ is a short hand notation for the sequence $X_1,\ldots,X_t$ and $Z^{t-1}$ and $\tilde{Y}^{t-1}$ are similarly defined.

This encoded symbol is transmitted over the forward channel (1a) producing a channel output $Y_t$. At the next time instant, $Y_t$ gets transmitted over the backward channel (1b).

The receiver consists of a decoder and a memory. The content of the memory is denoted by $M_t$ and takes values in a finite alphabet $\mathcal{M}$. At each stage $t$, the receiver generates an estimate $\hat{X}_t$ of the source taking values in a finite alphabet $\hat{\mathcal{X}}$ using a decoding rule $g_t$, i.e.,

$$\hat{X}_1 = g_1(Y_1), \tag{3a}$$
$$\hat{X}_t = g_t(Y_t, M_{t-1}), \qquad t = 2,\ldots,T, \tag{3b}$$

and updates the content of its memory using a memory update rule $l_t$, i.e.,

$$M_1 = l_1(Y_1), \tag{4a}$$
$$M_t = l_t(Y_t, M_{t-1}), \qquad t = 2,\ldots,T. \tag{4b}$$

The performance of the system is quantified by a uniformly bounded distortion function $\rho : \mathcal{X} \times \hat{\mathcal{X}} \to [0, \rho_{\max}]$, where $\rho_{\max} < \infty$. The distortion at time $t$ is given by $\rho(X_t, \hat{X}_t)$.

The collection $C := (c_1,\ldots,c_T)$ of encoding rules for the entire horizon is called an *encoding strategy*. Similarly, the collection $G := (g_1,\ldots,g_T)$ of decoding rules is called a *decoding strategy* and the collection $L := (l_1,\ldots,l_T)$ of memory update rules is called a *memory update strategy*. Further, the choice $(C,G,L)$ of communication rules for the entire horizon is called a *communication strategy* or a *design*.

The performance of a communication strategy is quantified by the expected total distortion under that strategy and is given by

$$\mathcal{J}_T(C, G, L) := \mathbb{E}\left\{\sum_{t=1}^{T} \rho(X_t, \hat{X}_t) \,\bigg|\, C, G, L\right\}. \quad (5)$$

We are interested in the following optimization problem:

*Problem 1:* Assume that the encoder and the receiver know the source statistics $P_{X_1}$ and $P_{X_{t+1}|X_t}$, $t = 1, \ldots, T$, the forward and backward channel functions $h$, $\tilde{h}$, the forward and the backward channel noise statistics $P_N$ and $P_{\tilde{N}}$, the distortion functions $\rho$ and the time horizon $T$. Choose a communication strategy $(C^*, G^*, L^*)$ that is optimal with respect to performance criterion of (5), i.e.,

$$\mathcal{J}_T(C^*, G^*, L^*) = \mathcal{J}_T^* := \min_{\substack{C \in \mathscr{C}^T \\ G \in \mathscr{G}^T \\ L \in \mathscr{L}^T}} \mathcal{J}_T(C, G, L), \quad (6)$$

where $\mathscr{C}^T := \mathscr{C}_1 \times \cdots \times \mathscr{C}_T$, $\mathscr{C}_t$ is the family of functions from $\mathcal{X}^t \times \tilde{\mathcal{Y}}^{t-1} \times \tilde{\mathcal{Z}}^{t-1}$ to $\mathcal{Z}$, $\mathscr{G}^T := \mathscr{G} \times \ldots \times \mathscr{G}$ ($T$–times), $\mathscr{G}$ is the family of functions from $\mathcal{Y} \times \mathcal{M}$ to $\hat{\mathcal{X}}$, $\mathscr{L}^T := \mathscr{L} \times \ldots \mathscr{L}$ ($T$–times), and $\mathscr{L}$ is the family of functions from $\mathcal{Y} \times \mathcal{M}$ to $\mathcal{M}$.

The design of an optimal communication strategy for a real–time communication system with noisy feedback has not been considered in the literature so far. The work on real–time communication assumes either a noiseless forward channel [1]–[4], or no feedback [5]–[8], or it assumes noiseless feedback [9]. The work on noisy feedback [10]–[13] does not assume a real–time constraint on information transmission.

The key contribution of this paper is the presentation of a systematic methodology for the design of globally optimal strategies for real–time communication with noisy feedback. We treat the design of an optimal communication strategy as a decentralized multi–agent sequential optimization problem. We show that an optimal communication strategy can be obtained by proceeding backwards in time and solving a set of nested optimality equations.

The rest of this paper is organized as follows. We present some preliminary results in Section II. Then we present qualitative properties of optimal encoding and decoding strategies in Section III and describe an algorithm for determining globally optimal communication strategies in Section IV. We conclude in Section V.

## II. SOME PRELIMINARIES

### A. Problem Classification

Problem 1 is a sequential stochastic optimization problem as defined in [14]. To understand the sequential nature of the problem, we need to refine the notion of time. We call each step of the system a *stage*. For each stage, we consider three time instances:[1] $t^+$, $t + 1/2$, and $(t+1)^-$. For the ease of notation, we will denote these time instances by $^1t$, $^2t$, and $^3t$, respectively. Assume that the system has three "agents", the encoder (agent 1), the decoder (agent 2), and the memory update (agent 3), which act sequentially at $^1t$, $^2t$, and $^3t$, respectively. The order in which the random variables are generated in the system is illustrated in Figure 2. Since the ordering of the decision makers can be done independently of the realization of the system variables, Property C of [15] is trivially satisfied and hence Problem 1 is a *causal sequential stochastic optimization* problem as defined in [14].

Problem 1 is a multi–agent problem where all agents have the same objective given by (6). Such problems are called team problems [16], and are further classified as static teams or dynamic teams on the basis of their information structure. In static teams, an agent's information is a function of primitive random variables only, while in dynamic teams, in general, an agent's information depends on the functional form of the decision rules of other agents. In Problem 1 the receiver's information depends on the functional form of the encoding rule. Thus Problem 1 is a dynamic team. Dynamic teams are, in general, functional optimization problems having a complex interdependence among the decision rules [17]. This interdependence leads to non–convex (in policy space) optimization problems that are hard to solve.

For the ease of notation, at time instances $^1t$, $^2t$, and $^3t$, we will denote the current decision rule by $^1\phi_t$, $^2\phi_t$, and $^3\phi_t$ and the past decision rules by $^1\phi^{t-1}$, $^2\phi^{t-1}$, and $^3\phi^{t-1}$, i.e.,

$$^1\phi_t := c_t, \qquad ^1\phi^{t-1} := (c^{t-1}, g^{t-1}, l^{t-1}), \quad (7a)$$
$$^2\phi_t := g_t, \qquad ^2\phi^{t-1} := (c^t, g^{t-1}, l^{t-1}), \quad (7b)$$
$$^3\phi_t := l_t. \qquad ^3\phi^{t-1} := (c^t, g^t, l^{t-1}). \quad (7c)$$

### B. The Notion of Information

We believe that the traditional information theoretic notions entropy and mutual information are asymptotic concepts which are not directly applicable to real–time communication problems. So, we first describe a decision theoretic notion of information. Let $(\Omega, \mathfrak{F}, P)$ be the probability field with respect to which all primitive random variables are defined. Suppose $^iO_t$ is the observation of agent $i$ at time $^it$, and $^i\phi^{t-1}$ is the past decision rules of all agents. Since the problem is sequential, for any choice of $^i\phi^{t-1}$, $^iO_t$ is measurable with respect to $\mathfrak{F}$. Furthermore, for any choice of $^i\phi^{t-1}$, let $\sigma(^iO_t; ^i\phi^{t-1})$ denote the smallest subfield of $\mathfrak{F}$ with respect to which $^iO_t$ is measurable. Then, the *information field* of agent $k$ at time $^it$ is $\sigma(^iO; ^i\phi^{t-1})$. Using this notion of information, we define variables that represent the information field at the encoder's and receiver's sites just before each agent acts on the system.

*Definition 1:* Let $^1E_t$, $^2E_t$, and $^3E_t$ denote the observation and $^1\mathfrak{E}_t$, $^2\mathfrak{E}_t$, and $^3\mathfrak{E}_t$ denote the *information field* at the encoder's site at time $^1t$, $^2t$, and $^3t$, respectively, i.e.,

$$^1E_t := (X^t, Z^{t-1}, \tilde{Y}^{t-1}), \quad ^1\mathfrak{E}_t := \sigma(^1E_t; ^1\phi^{t-1}), \quad (8a)$$

---
[1]The actual values of these time instances is not important; we just need three values in increasing order.

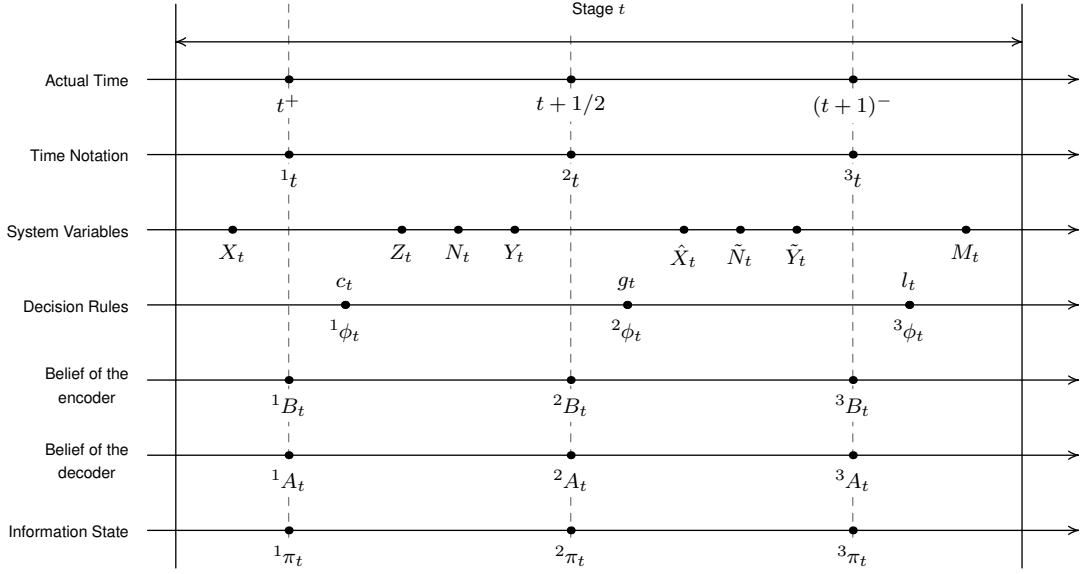

Fig 2. Sequential ordering of different variables in the system

$$^2E_t := (X^t, Z^t, \tilde{Y}^{t-1}), \qquad {}^2\mathfrak{E}_t := \sigma(^2E_t; {}^2\phi^{t-1}), \quad (8b)$$
$$^3E_t := (X^t, Z^t, \tilde{Y}^t), \qquad {}^3\mathfrak{E}_t := \sigma(^3E_t; {}^3\phi^{t-1}). \quad (8c)$$

Further, let $^1R_t$, $^2R_t$, and $^3R_t$ denote the observation and $^1\mathfrak{R}_t$, $^2\mathfrak{R}_t$, and $^3\mathfrak{R}_t$ denote the *information field* at the receiver's site at time $^1t$, $^2t$, and $^3t$, respectively, i.e.,

$$^1R_t := (M_{t-1}), \qquad {}^1\mathfrak{R}_t := \sigma(^1R_t; {}^1\phi^{t-1}), \quad (9a)$$
$$^2R_t := (Y_t, M_{t-1}), \qquad {}^2\mathfrak{R}_t := \sigma(^2R_t; {}^2\phi^{t-1}), \quad (9b)$$
$$^3R_t := (Y_t, M_{t-1}), \qquad {}^3\mathfrak{R}_t := \sigma(^3R_t; {}^3\phi^{t-1}). \quad (9c)$$

Problem 1 is a decentralized problem because, at any time $t$, the information fields at the encoder's site and the receiver's site are non–comparable, that is, $^1\mathfrak{E}_t \not\subseteq {}^1\mathfrak{R}_t$ and $^1\mathfrak{E}_t \not\supseteq {}^1\mathfrak{R}_t$; and similar relations hold between $^2\mathfrak{E}_t$ and $^2\mathfrak{R}_t$, and between $^3\mathfrak{E}_t$ and $^3\mathfrak{R}_t$. Thus, at no time during the evolution of the system does the encoder "know" exactly what is "known" to the receiver and vice–versa. *Hence the information in the system is decentralized*. Notice that the information fields at the encoder and the receiver are coupled through decision rules. $^1\mathfrak{E}_1$ and $^1\mathfrak{R}_1$ are known before the system starts operating. The choice of $^1\phi_1$ determines $^2\mathfrak{E}_1$ and $^1\mathfrak{R}_1$, the choice of $^2\phi_1$ determines $^3\mathfrak{E}_1$ and $^3\mathfrak{R}_1$, and so on. Thus, $^1\mathfrak{E}_t$ and $^1\mathfrak{R}_t$ are determined completely by $^1\mathfrak{E}_1$, $^1\phi^{t-1}$ and $^1\mathfrak{R}_1$, $^1\phi^{t-1}$, respectively. Thus, the information $^1\mathfrak{E}_t$ and $^1\mathfrak{R}_t$ is coupled through the past decision rules $^1\phi^{t-1}$. Hence, Problem 1 has a *non–classical information structure* (see [18, 19]).

### C. Agent's Beliefs and their Evolution

Due to decentralization of information, it is important to characterize what one agent thinks about the other agent's observation, i.e., what the encoder "thinks" that the receiver "sees" and what the receiver "thinks" that the encoder "sees". This is captured by the encoder's belief about the observations of the receiver, and the receiver's belief about the observations of the encoder at time instances $^1t$, $^2t$, and $^3t$. These beliefs are given below.

*Definition 2:* Let $^1B_t$, $^2B_t$, and $^3B_t$ denote the encoder's belief about the receiver's observation at $^1t$, $^2t$, and $^3t$, respectively, i.e., for $i = 1, 2, 3$,

$$^iB_t(^ir) := \Pr\left(^iR_t = {}^ir \,\middle|\, {}^i\mathfrak{E}_t\right). \quad (10)$$

*Definition 3:* Let $^1A_t$, $^2A_t$, and $^3A_t$ denote the receiver's belief about the encoder's observation at $^1t$, $^2t$, and $^3t$, respectively, i.e., for $i = 1, 2, 3$,

$$^iA_t(^ie) := \Pr\left(^iE_t = {}^ie \,\middle|\, {}^i\mathfrak{R}_t\right). \quad (11a)$$

Further, let $\hat{A}_t$ denote the receiver's belief about the source output at time instance $^2t$, i.e.,

$$\hat{A}_t(x_t) := \Pr\left(X_t = x_t \,\middle|\, {}^2\mathfrak{R}_t\right). \quad (11b)$$

The sequential ordering of these beliefs is shown in Figure 2. For any particular realization $^1e_t$ of $^1E_t$, and any arbitrary (but fixed) choice of $^1\phi^{t-1}$, the realization $^1b_t$ of $^1B_t$ is a PMF on $\mathcal{M}$. If $E_t$ is a random vector, then $^1B_t$ is a random vector belonging to $\mathbb{P}(\mathcal{M})$, the space of PMFs on $\mathcal{M}$. Similar interpretations hold for $^2B_t$, $^3B_t$, $^1A_t$, $^2A_t$, and $^3A_t$.

The time evolution of these beliefs of the encoder and the receiver are coupled through their decision rules. Specifically,

*Lemma 1:* For each stage $t$, there exist deterministic functions $^1F$, $^2F$, and $^3F$ such that

$$^1B_t = {}^1F(^3B_{t-1}, l_{t-1}), \quad (12a)$$
$$^2B_t = {}^2F(^1B_t, Z_t), \quad (12b)$$
$$^3B_t = {}^3F(^2B_t, \tilde{Y}_t). \quad (12c)$$

The functions $^1F$ and $^2F$ are linear in their first argument.

*Lemma 2:* For each stage $t$, there exist deterministic functions $^2K$, $^3K$ and $\hat{K}$ such that

$$^2A_t = {}^2K(^1A_t, Y_t, c_t), \quad (13a)$$

$$^3A_t = {}^3K(^2A_t), \quad (13b)$$

$$\hat{A}_t = \hat{K}(^2A_t). \quad (13c)$$

The functions $^3K$ and $\hat{K}$ are linear in their first argument. Further, there exist deterministic functions $^1K_t$ for each $t$ such that

$$^1A_t = {}^1K_t(^1A_1, M_{t-1}, c^{t-1}, l^{t-1}). \quad (13d)$$

Due to lack of space the complete proofs of these lemmas are omitted. Detailed proofs can be found in [20].

An observation that simplifies the global optimization problem is the fact that the beliefs $^1B_t$ and $^1A_t$ are independent of the decoding strategy $G$. This is because decoding is a filtering problem that does not affect the future evolution of the system.

Before looking at the global optimization problem, we first identify qualitative properties of optimal encoders and decoders.

### III. Structural Results

In this section, we provide qualitative properties of optimal encoders (respectively, decoders) that are true for all arbitrary but fixed decoding and memory update strategies (respectively, encoding and memory update strategies).

#### A. Structural Results of Optimal Real–Time Encoders

*Theorem 1:* Consider Problem 1 for any arbitrary (but fixed) decoding and memory update strategies, $G = (g_1, \ldots, g_T)$ and $L = (l_1, \ldots, l_T)$, respectively. Then there is no loss in optimality in restricting attention to encoding rules of the form

$$Z_t = c_t(X_t, {}^1B_t), \quad t = 2, \ldots, T. \quad (14)$$

*Proof.* We look at the problem from the encoder's point of view. Note that $\{X_t, t = 1, \ldots, T\}$ is a Markov process independent of the noise in the forward and the backward channel. This fact together with the results of Lemma 1 implies that

$$\Pr\left(X_{t+1}, {}^1B_{t+1} \mid X^t, {}^1B^t, Z^t, c^t, g^t, l^t\right)$$
$$= \Pr(X_{t+1} \mid X_t) \Pr\left({}^1B_{t+1} \mid {}^1B_t, Z_t, l_t\right)$$
$$= \Pr\left(X_{t+1}, {}^1B_{t+1} \mid X_t, {}^1B_t, Z_t, l_t\right) \quad (15)$$

Thus $\{(X_t, {}^1B_t), t = 1, \ldots, T\}$ is a controlled Markov process with control action $Z_t$. Further, the expected conditional instantaneous distortion can be written as

$$\mathbb{E}\left\{\rho(X_t, \hat{X}_t) \mid {}^3\mathfrak{E}_t\right\} =$$

$$= \sum_{\substack{y_t \in \mathcal{Y} \\ m_{t-1} \in \mathcal{M}}} \rho(X_t, g_t(y_t, m_{t-1})) \Pr\left(y_t, m_{t-1} \mid {}^3\mathfrak{E}_t\right)$$

$$= \sum_{\substack{y_t \in \mathcal{Y} \\ m_{t-1} \in \mathcal{M}}} \rho(X_t, g_t(y_t, m_{t-1})) \, {}^2F(^1B_t, Z_t)$$

$$=: {}^1\rho(X_t, {}^1B_t, Z_t, g_t). \quad (16)$$

Thus, the total expected distortion can be written as

$$\mathbb{E}\left\{\sum_{t=1}^T \rho(X_t, \hat{X}_t) \,\bigg|\, C, G, L\right\}$$
$$= \mathbb{E}\left\{\sum_{t=1}^T \mathbb{E}\left\{\rho(X_t, \hat{X}_t) \,\bigg|\, {}^3\mathfrak{E}_t\right\} \,\bigg|\, C, G, L\right\}$$
$$= \mathbb{E}\left\{\sum_{t=1}^T {}^1\rho(X_t, {}^1B_t, Z_t, g_t) \,\bigg|\, C, G, L\right\}. \quad (17)$$

Hence from the encoder's point of view, we have a perfectly observed controlled Markov process $\{(X_t, {}^1B_t), t = 1, \ldots, T\}$ with control action $Z_t$ and an instantaneous distortion $^1\rho(X_t, {}^1B_t, Z_t, g_t)$ (recall that $G$ is fixed). From Markov decision theory [21] we know that there is no loss of optimality in restricting attention to encoding rules of the form (14). □

Theorem 1 immediately implies the following:

*Corollary 1:* The optimal performance $\mathcal{J}_T^*$ given by (6) can be determined by

$$\mathcal{J}_T(C^*, G^*, L^*) = \mathcal{J}_T^* := \min_{\substack{C \in \mathscr{C}_S^T \\ G \in \mathscr{G}^T \\ L \in \mathscr{L}^T}} \mathcal{J}_T(C, G, L), \quad (18)$$

where $\mathscr{C}_S^T := \mathscr{C}_S \times \cdots \times \mathscr{C}_S$ ($T$–times), $\mathscr{C}_S{}^2$ is the space of functions from $\mathcal{X} \times \mathbb{P}(\mathcal{M})$ to $\mathcal{Z}$, $\mathscr{G}^T$ and $\mathscr{L}^T$ are defined as before, and $\mathbb{P}(\mathcal{M})$ denotes the space of all probability measures on $\mathcal{M}$.

#### B. Structure of Optimal Real–Time Decoders

*Theorem 2:* Consider Problem 1 for any arbitrary (but fixed) encoding and memory update strategies, $C = (c_1, \ldots, c_T)$ and $L = (l_1, \ldots, l_T)$, respectively. Then there is no loss in optimality in restricting attention to decoding rules of the form

$$\hat{X}_t = \hat{g}(\hat{A}_t) := \arg\min_{\hat{x} \in \hat{\mathcal{X}}} \sum_{x \in \mathcal{X}} \rho(x, \hat{x}) \hat{A}_t(x). \quad (19)$$

*Proof.* We look at the problem from the decoder's point of view. Since decoding is a filtering problem, minimizing the total distortion $\mathcal{J}_T(C, G, L)$ is equivalent to minimizing the conditional expected instantaneous distortion $\mathbb{E}\left\{\rho(X_t, \hat{X}_t) \mid {}^2\mathfrak{R}_t\right\}$ for each time $t$. This conditional expected instantaneous distortion can be written as

$$\mathbb{E}\left\{\rho(X_t,\hat{X}_t)\,\Big|\,{}^2\mathfrak{R}_t\right\} = \sum_{x_t \in \mathcal{X}} \rho(x_t,\hat{X}_t)\Pr\left(x_t\,\big|\,{}^2\mathfrak{R}_t\right)$$
$$= \sum_{x_t \in \mathcal{X}} \rho(x_t,\hat{X}_t)\hat{A}_t(x_t)$$

and is minimized by the decoding rule given in (19). □

---
[2]Note that the $S$ in $\mathscr{C}_S$ is a short form of separated, and should not be confused with $\mathscr{C}_t$ defined in Problem 1.

## IV. Determining Globally Optimal Communication Strategy

A globally optimal design for Problem 1 always exists because there are finitely many designs and we can always choose the one with best performance. However, a brute force evaluation of each design to find the optimal one is computationally impractical. So we want to determine a systematic algorithm to search for an optimal design. One such systematic approach, called *sequential decomposition*, is to sequentially determine optimal decision rules for all stages by proceeding backwards in time. This procedure simplifies exponentially the complexity of searching for an optimal solution. The resultant "simplified" problem is still exponential in complexity, which reflects the complexity of finding optimal strategies in decentralized systems.

The key step in obtaining a sequential decomposition is to identify an *information state sufficient for performance evaluation* (also called a *sufficient statistic for control*). For Problem 1 one such information state is given by the following unconditional probability laws.

*Definition 4:* Define ${}^1\pi_t$, ${}^2\pi_t$, and ${}^3\pi_t$ as follows:
$$ {}^1\pi_t = \Pr\left(X_t, M_{t-1}, {}^1B_t\right), \quad (20a)$$
$$ {}^2\pi_t = \Pr\left(X_t, Y_t, M_{t-1}, {}^2B_t\right), \quad (20b)$$
$$ {}^3\pi_t = \Pr\left(X_t, Y_t, \tilde{Y}_t, M_{t-1}, {}^3B_t\right). \quad (20c)$$

Let ${}^1\Pi$ denote the space of probability measures on $\mathcal{X} \times \mathcal{M} \times \mathbb{P}(\mathcal{M})$, ${}^2\Pi$ denote the space of probability measures on $\mathcal{X} \times \mathcal{Y} \times \mathcal{M} \times \mathbb{P}(\mathcal{Y} \times \mathcal{M})$ and ${}^3\Pi$ denote the space of probability measures on $\mathcal{X} \times \mathcal{Y} \times \tilde{\mathcal{Y}} \times \mathcal{M} \times \mathbb{P}(\mathcal{Y} \times \mathcal{M})$. Then, ${}^1\pi_t$ takes values in ${}^1\Pi$, ${}^2\pi_t$ takes values in ${}^2\Pi$ and ${}^3\pi_t$ takes values in ${}^3\Pi$.

*Lemma 3:* ${}^1\pi_t$, ${}^2\pi_t$, and ${}^3\pi_t$ are information states for the encoder, decoder, and memory update, respectively, i.e.,
1. there are linear transformations ${}^1Q$, ${}^2Q$, and ${}^3Q$ such that
$$ {}^2\pi_t = {}^1Q(c_t)\,{}^1\pi_t, \quad (21a)$$
$$ {}^3\pi_t = {}^2Q\,{}^2\pi_t, \quad (21b)$$
$$ {}^1\pi_{t+1} = {}^3Q(l_t)\,{}^3\pi_t. \quad (21c)$$

2. the expected instantaneous cost can be expressed as
$$ \mathbb{E}\left\{\rho(X_t,\hat{X}_t)\,\Big|\,c^t,g^t,l^{t-1}\right\} = {}^2\rho({}^2\pi_t, g_t) \quad (22)$$

where ${}^2\rho(\cdot)$ is a deterministic function.

Due to lack of space, the proof of this Lemma is omitted. Detailed proof can be found in [20]. Using this result, the performance criterion of (5) can be rewritten as
$$\mathcal{J}_T(C,G,L) = \mathbb{E}\left\{\sum_{t=1}^T \rho(X_t,\hat{X}_t)\,\bigg|\,C,G,L\right\}$$
$$\stackrel{(a)}{=} \sum_{t=1}^T \mathbb{E}\left\{\rho(X_t,\hat{X}_t)\,\Big|\,c^t,g^t,l^{t-1}\right\}$$
$$\stackrel{(b)}{=} \sum_{t=1}^T {}^2\rho({}^2\pi_t, g_t) \quad (23)$$

where $(a)$ follows from the sequential ordering of system variables and $(b)$ follows from Lemma 3.

### A. An Equivalent Optimization Problem

Consider a centralized deterministic optimization problem with state space alternating between ${}^1\Pi$, ${}^2\Pi$, and ${}^3\Pi$ and action space alternating between $\mathscr{C}_S$, $\mathscr{G}$, and $\mathscr{L}$. The system dynamics are given by (21) and at each stage $t$ the decision rules $c_t$, $g_t$ and $l_t$ are determined according to *meta–rules* ${}^1\Delta_t$, ${}^2\Delta_t$, and ${}^3\Delta_t$, where ${}^1\Delta_t$ is a function from ${}^1\Pi$ to $\mathscr{C}_S$, ${}^2\Delta_t$ is a function from ${}^2\Pi$ to $\mathscr{G}$ and ${}^3\Delta_t$ is a function from ${}^3\Pi$ to $\mathscr{L}$. Thus the system equations (21) can be written as
$$ c_t = {}^1\Delta_t({}^1\pi_t), \quad {}^2\pi_t = {}^1Q(c_t)\,{}^1\pi_t, \quad (24a)$$
$$ g_t = {}^2\Delta_t({}^2\pi_t), \quad {}^3\pi_t = {}^2Q\,{}^2\pi_t, \quad (24b)$$
$$ l_t = {}^3\Delta_t({}^3\pi_t), \quad {}^1\pi_{t+1} = {}^3Q(l_t)\,{}^3\pi_t. \quad (24c)$$

At each stage an instantaneous cost ${}^2\rho({}^2\pi_t, g_t)$ is incurred. The choice $({}^1\Delta_1, {}^2\Delta_1, {}^3\Delta_1, \ldots, , {}^1\Delta_T, {}^2\Delta_T, {}^3\Delta_T)$ is called a *meta–design* and denoted by $\Delta^T$. The performance of a meta–design is given by the total cost incurred by that meta–design, i.e.,
$$ \mathcal{J}_T(\Delta^T) = \sum_{t=1}^T {}^2\rho({}^2\pi_t, g_t). \quad (25)$$

Now consider the following optimization problem:

*Problem 2:* Consider the dynamic system (24) with known transformations ${}^1Q$, ${}^2Q$, and ${}^3Q$. The initial state ${}^1\pi_1$ is given. Determine a meta–design $\Delta^T$ to minimize the total cost given by (25).

Observe that for any initial state ${}^1\pi_1$, a choice of meta–design $\Delta^T$ determines a design $(C,G,L)$ through (24). Relation (22) implies that the expected distortion under design $(C,G,L)$, given by (5), is equal to the cost under meta–design $\Delta^T$ given by (25). Thus, if the transformation ${}^1Q$, ${}^2Q$, and ${}^3Q$ in Problem 2 are chosen as in Lemma 3, an optimal meta–design for Problem 2 determines an optimal design for Problem 1. Problem 2 is a classical deterministic control problem and optimal meta–designs can be determined as follows:

*Theorem 3:* An optimal meta–design $\Delta^{*,T}$ for Problem 2, and consequently an optimal design $(C^*, G^*, L^*)$ for Problem 1 can be determined as follows. For any ${}^1\pi \in {}^1\Pi$, ${}^2\pi \in {}^2\Pi$, and ${}^3\pi \in {}^3\Pi$, define the following functions:

$$^1V_{T+1}(^1\pi) = 0, \tag{26a}$$

and for $t = 1, \ldots, T$

$$^1V_t(^1\pi) = \inf_{c \in \mathscr{C}_S} {}^2V_t\bigl(^1Q(c)\,^1\pi\bigr), \tag{26b}$$

$$^2V_t(^2\pi) = \min_{g \in \mathscr{G}} {}^2\rho(^2\pi, g) + {}^3V_t(^2Q\,^2\pi), \tag{26c}$$

$$^3V_t(^3\pi) = \min_{l \in \mathscr{L}} {}^1V_{t+1}\bigl(^3Q(l)\,^3\pi\bigr). \tag{26d}$$

The arg min (or arg inf) at each step determines the optimal meta–design $\Delta^{*,T}$. After an optimal meta–design has been determined, an optimal design $(C^*, G^*, L^*)$ can be determined through (24). Furthermore, the optimal performance is given by

$$\mathcal{J}_T^* = {}^1V_1(^1\pi_1). \tag{27}$$

*Proof.* This is a standard result, see [21, Chapter 2].
□

The above nested optimality equations determine a globally optimal design and the globally optimal performance. Observe that the functional form of the optimality equations does not change with time. So, the results presented here can be easily extended to infinite horizon problem with expected discounted distortion or average distortion per unit time criteria. Such an extension to infinite horizon will result in a fixed point equation to determine a time–invariant (stationary) meta–design; *the design at each stage will be time varying*. Due to decentralization of information, optimal designs for infinite horizon are not stationary. This phenomenon also occurs in real–time communication with no feedback.

## V. Discussion and Conclusion

The solution framework presented in this paper and some of our previous papers [7, 8, 20] provides an alternative approach to real–time communication problems. Information theoretic performance bounds or coding theoretic low–complexity coding schemes are not known for noisy real–time communication systems. In the absence of such results, the designer of a real–time communication system has to choose a good heuristic communication strategy and hope that it meets the performance requirements. If it does not, the designer needs to try different communication strategies until one that meets the performance requirements is found.

In this paper we have presented an alternative, systematic methodology to design an optimal communication strategy for real–time communication systems with noisy feedback. Instead of trying out heuristic strategies one by one, optimal communication strategies can be determined by solving the nested optimality equations of Theorem 3. Note that these are not typical dynamic programming equations as each step is a functional optimization problem. Hence, although the systematic methodology presented here exponentially simplifies the complexity of finding an optimal design as compared to a brute force approach, solving the resultant nested optimality equations is a formidable computational task. It may be possible to extend the computational techniques for solving dynamic programming equations to efficiently solve equations of the form (26). The solution of (26) also determines the optimal performance of the system and can be used to check the degree of sub-optimality of heuristic designs.

### Acknowledgements

This research was supported in part by NSF Grant CCR-0325571 and NASA Grant NNX06AD47G. The authors are grateful to A. Anastasopoulos and S. Pradhan for insightful discussions. They are also grateful to the anonymous reviewers whose comments helped in improving the presentation of this paper.